\documentclass[aps,preprint,prl,showpacs]{revtex4}

\usepackage{graphicx}
\begin{document}



\title{Observation of Spin-Orbit Berry's Phase in Magnetoresistance of a Two-Dimensional Hole Anti-dot System}

\author{Ning Kang}\email{ning.kang@issp.u-tokyo.ac.jp}
\author{Eisuke Abe, Yoshiaki Hashimoto, Yasuhiro Iye, and Shingo
Katsumoto}
\affiliation{Institute for Solid State Physics, University of Tokyo, \\
5-1-5 Kashiwanoha, Kashiwa, Chiba 277-8581, Japan}

\date{\today}

\begin{abstract}
We report observation of spin-orbit Berry's phase in the
Aharonov-Bohm (AB) type oscillation of weak field
magnetoresistance in an anti-dot lattice (ADL) of a
two-dimensional hole system. An AB-type oscillation is superposed
on the commensurability peak, and the main peak in the Fourier
transform is clearly split up due to variation in Berry's phase
originating from the spin-orbit interaction. A simulation
considering Berry's phase and the phase arising from the
spin-orbit shift in the momentum space shows qualitative agreement
with the experiment.
\end{abstract}


\maketitle

\begin{figure}
\parbox{0.495\linewidth}{(a)
\includegraphics[width=\linewidth]{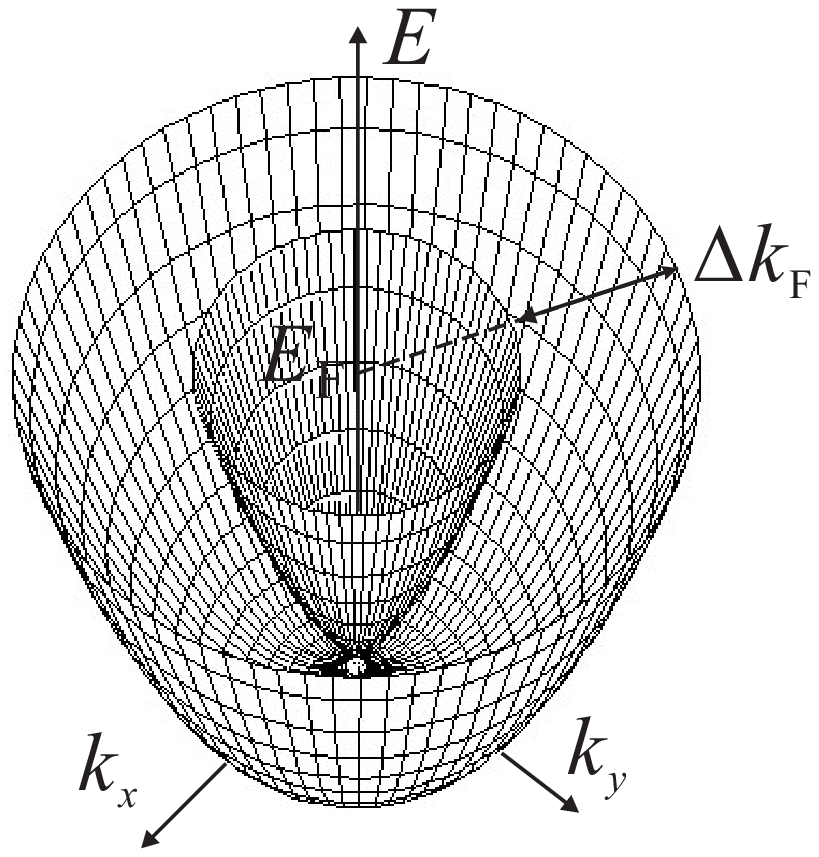}\\
(b)
\includegraphics[width=\linewidth]{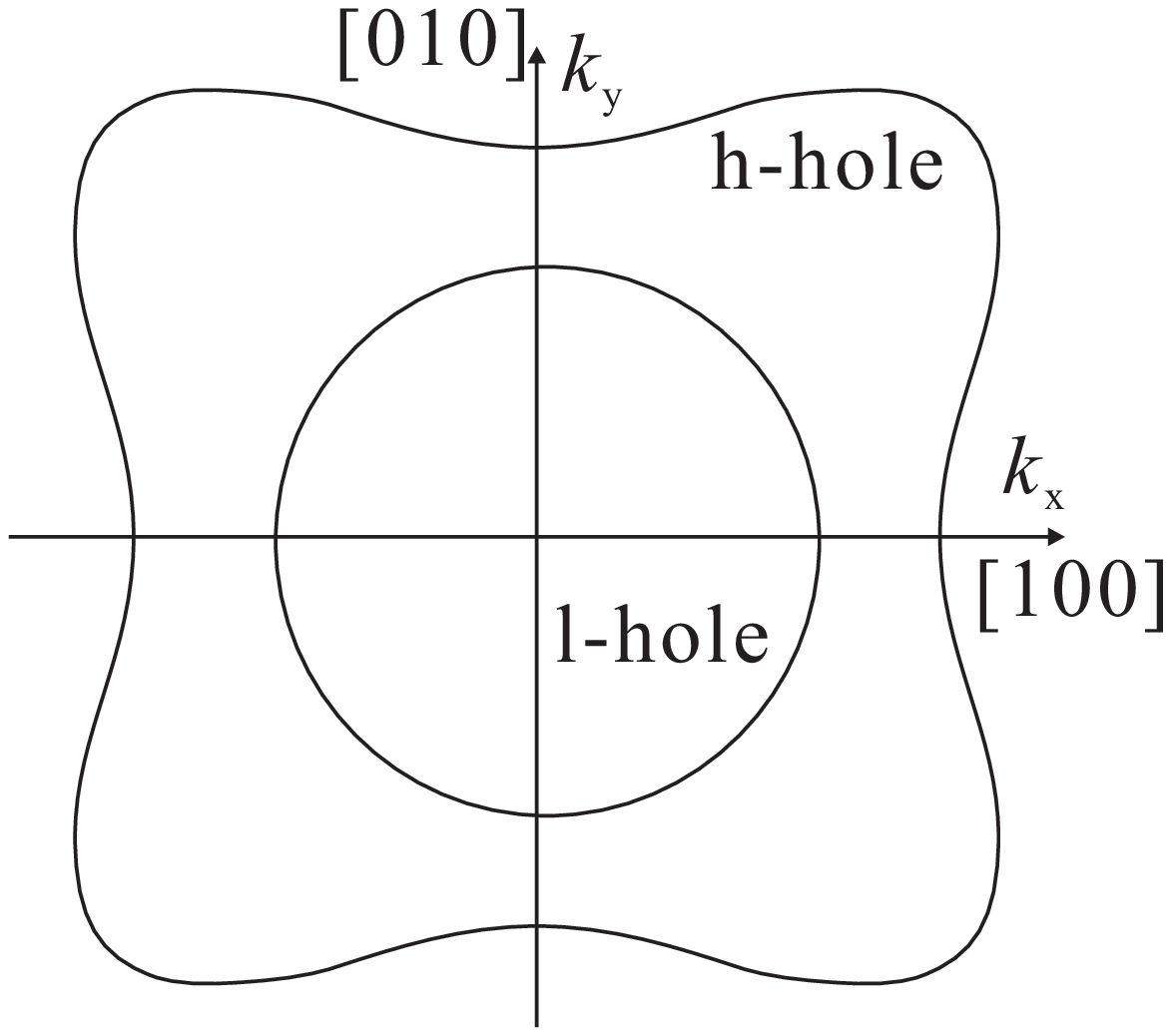}}
\parbox{0.495\linewidth}{(c)
\includegraphics[width=1.0\linewidth]{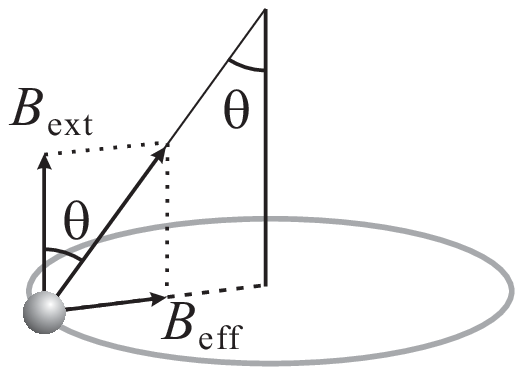}\\
\rule{0mm}{10mm}\\
(d)
\includegraphics[width=0.9\linewidth]{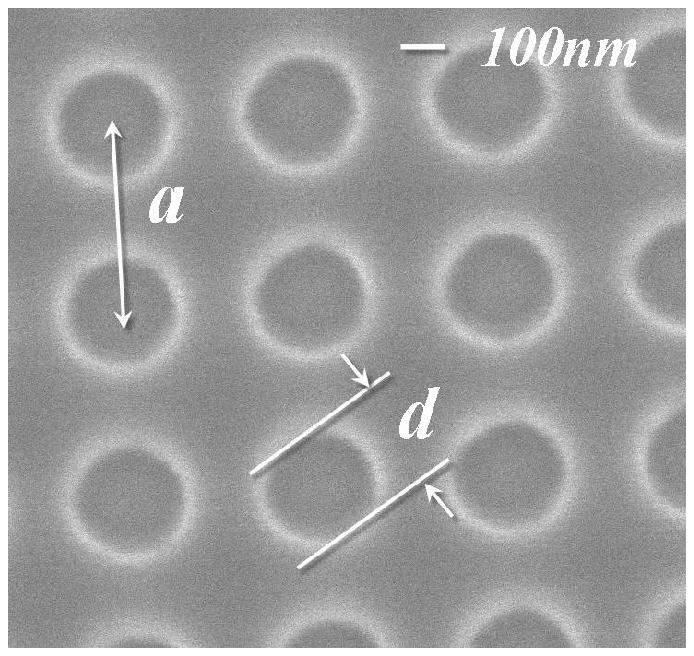}\\
}\caption{ (a) Surface plot of spin subbands explains two Fermi
contours with different $k_{\rm F}$ appear as a result of
spin-orbit interaction. (b) According to more realistic
calculation\cite{Ando2,Ekenberg}, the outer subband (h-hole) is
significantly warped. (c) Schematic illustration of $B_{\rm eff}$
seen by holes moving around a ring structure in the presence of
spin-orbit interaction. (d) Scanning electron micrograph of the
ADL sample. } \label{intro_fig}
\end{figure}

\vspace{2mm} Spin-orbit interaction in two-dimensional system is
predicted to introduce an additional geometric phase (Berry's
phase) \cite{Loss,Aronov,Stern,Qian,Meir}. Here we consider the
case of two-dimensional hole system (2DHS), in which the
spin-orbit interaction is strong in III-V compound semiconductors
due to the $p$-orbital nature of the valence bands. In a 2DHS with
the Rashba-type spin-orbit coupling, the additional phase shifts
arise in two ways. The spin-orbit interaction causes split of the
parabolic dispersion, producing two Fermi circles of opposite
spins (Fig.\ref{intro_fig}(a),(b))
\cite{Ando2,Ekenberg,Eisenstein,Lu,Iye2}. This means, the hole
system has two different Fermi wave vectors, of which the
difference ($\Delta k_{\rm F}$) gives a phase shift and modifies
the interference through, {\it e.g.}, a ring structure. Another
shift in phase comes from the spin part of the wavefunction. When
a hole moves along a closed contour in the real space, the same
occurs in $k$-space and it feels effective magnetic field $B_{\rm
eff}$ due to the spin-orbit interaction. $B_{\rm eff}$ is in the
2D plane and orthogonal to the velocity. Hence when $B_{\rm eff}$
is strong enough, the spin placed in the 2D plane turns by 2$\pi$
giving Berry's phase of 0 or $\pi$.

When an external magnetic field $B_{\rm ext}$ is applied
perpendicular to the 2DHS plane, the hole moving along a loop
acquires an Aharonov-Bohm (AB) phase. At the same time, the spin
gets some component perpendicular to the plane, which means some
diminishing in the in-plane component (Fig.1(c)). This hence affects the
above two additional phases. The variation in Berry's phase
($\Delta\theta_{\rm B}$) and that in the phase due to $\Delta
k_{\rm F}$ ($\Delta\theta_k$) are given as\cite{Loss,Aronov,Stern,Qian,Meir,Yau}
\begin{equation}
\Delta\theta_{\rm B} = \pi(1-\cos\theta),\label{ph1}
\end{equation}
\begin{equation}
\Delta\theta_k = \pi r \Delta k_{\rm F} \sin\theta, \label{ph2}
\end{equation}
where $\theta\equiv\arctan(B_{\rm eff}/B_{\rm ext})$, and $r$ is
the radius of the AB ring. Though significant warping of the outer
Fermi contour is reported\cite{Ando2,Ekenberg} as shown in
Fig.1(b), we can consider $\Delta k_{\rm F}$ as an average over
the perimeter in the present discussion. Hence the phase of an AB
oscillation in a ring made of such a 2DHG should show continuous
shift giving a nontrivial peak shape in its Fourier transform (FT)
amplitude versus the frequency when the FT is performed in a
finite field range. In experiments, such phases have been observed
as splitting of AB magnetoresistance oscillation frequency in
single AB rings\cite{Mopurgo,Yau,Yang}. Especially the authors of
Ref. 12 observed nontrivial variation of the main FT peak due to
the phase shift represented in Eqs.\ref{ph1} and \ref{ph2}.

In an anti-dot structure, the carriers are excluded from a dot
region and magnetoresistance oscillation similar to that in an AB
ring is expected. In an anti-dot lattice (ADL), AB oscillations of
paths around single anti-dots should be averaged out because of
random phasing, whereas the effect of the AB phase remains in the
density of states resulting in magnetoresistance oscillation
called AB-type oscillation\cite{Weiss,Ando,Nihey,Iye}. It is,
then, of strong interest whether the effect of spin-orbit Berry's
phase appears in the density of states responsible for the AB-type
oscillation.

In this letter, we report observation of the predicted Berry's
phase through clear splitting of the main peak in FT of the
AB-type oscillation in the magnetoresistance of an ADL. The shift
in $k$-space is independently estimated from the Shubnikov-de Haas (SdH)
oscillation.
Because the ADL automatically performs
the disorder average, random specific
effects in $h/e$ oscillation are suppressed
and intrinsic information can be obtained.

An (Al,Ga)As/GaAs heterostructure was grown by molecular beam
epitaxy on a (001) GaAs substrate. Delta-doped Be layers produced
a 2DHS with the hole concentration of
2.3$\times$10$^{11}$cm$^{-2}$ and the mobility of
6.8$\times$10$^4$ cm$^{2}$/Vs at 4.2K. SdH oscillation in the
magnetoresistance of the 2DHS without ADL was measured at 60mK.
From the plot of the peak index versus the inverse magnetic field,
the concentrations of holes with smaller Fermi contour and larger
one as 0.78$\times$10$^{11}$cm$^{-2}$ and
1.56$\times$10$^{11}$cm$^{-2}$ respectively. In this letter, we
call the former and the latter as l-holes and h-holes,
respectively. In the simple effective mass model, these two
spin-split hole bands correspond to the two Fermi contours in
Fig.\ref{intro_fig}(a), though significant warping in h-hole band
is reported in more realistic band
calculation\cite{Ando2,Ekenberg}. The grown film was wet-etched
into Hall bars with [110] current direction. Circular anti-dot
shape is adopted and ADLs were defined by electron beam
lithography and wet-etching to a depth of 50nm. An ADL is
characterized mainly by the dot diameter $d$, the lattice
structure, and the lattice period $a$. In this study, we designed
a square lattice aligned along [110] ([$\bar{1}$10]) with
$d$=250nm and $a$=500nm (see Fig.\ref{intro_fig}(d)).

The sample was directly immersed in $^3$He-$^4$He mixture in
a mixing chamber of a dilution refrigerator and cooled down to 60mK.
$B_{\rm ext}$ up to 6T was applied by a superconducting solenoid.
The resistance was measured in four-terminal configuration
by standard lock-in technique with a frequency of 80Hz.

\begin{figure}
\includegraphics[width=0.8\linewidth, clip]{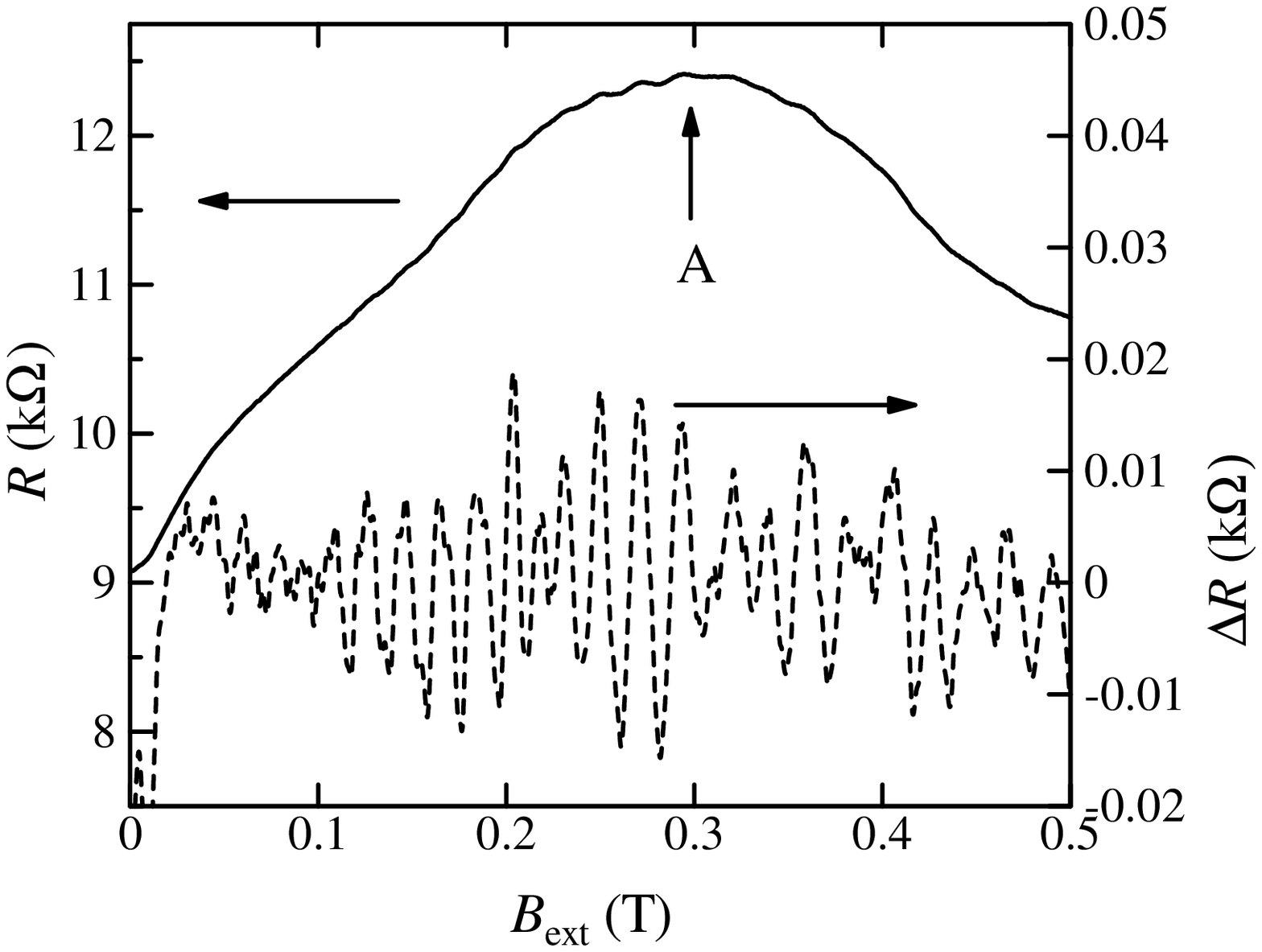}
\caption{ Upper solid line: Resistance of the sample as a function
of the external magnetic field $B_{\rm ext}$ for $T$=60 mK. Lower
broken line: Extracted oscillating part of the magnetoresistance.
For the background subtraction, a 12th order polynomial for each
set of 100 consecutive data points is adopted. } \label{rawR}
\end{figure}

The solid line in Fig.\ref{rawR} shows the resistance of the
sample as a function of $B_{\rm ext}$ up to 0.5T. In higher field
region, formation of the edge states and alignment of hole spins
perpendicular to the plane bring about different physics.
Therefore, we deal mainly with this low-field region. The
magnetoresistance shows a clear commensurability peak marked as A
by an arrow in the Fig.\ref{rawR}. Generally such peaks in
resistance appear when the carrier cyclotron orbits are
commensurate with an ADL and localized in the lattice. For a
circular Fermi contour, the classical cyclotron radius $R_c$ is
expressed as
\begin{equation}
R_c=\frac{\hbar k_{\rm F}}{eB_{\rm ext}}, \label{cyclo}
\end{equation}
where the Fermi wavevector $k_{\rm F}$ is written as $\sqrt{4\pi
p}$ with the hole concentration $p$ (here we assume the spin
degeneracy is lifted). $B_{\rm ext}$ at peak position A in
Fig.\ref{rawR} can be well ascribed to localized orbits of l-hole
encircling single anti-dot. For h-hole, as noted above, we should
take into account the effect of warped Fermi contour in
Fig.\ref{intro_fig}(b). This effect will be discussed in a
separate paper. For a small hump structure around 0.05T, we do not
have any appropriate interpretation at present.

As shown in Fig.\ref{rawR}, an oscillation with a short period
superposed on the commensurability peak is visible. In order to
extract the oscillating component, we subtract the background
commensurability structure by fitting slowly-varying functions. An
example of extracted resistance variation is shown with the broken
line in Fig.\ref{rawR}, which exhibits a clear fast oscillation.
Here for the background subtraction, a 12th order polynomial for
each set of 100 consecutive data points is adopted. Although the
selection of fitting function affects the outlook of the
oscillatory component, little is affected in the main and sub peak
structures in the Fourier space.

\begin{figure}
\includegraphics[width=0.8\linewidth,clip]{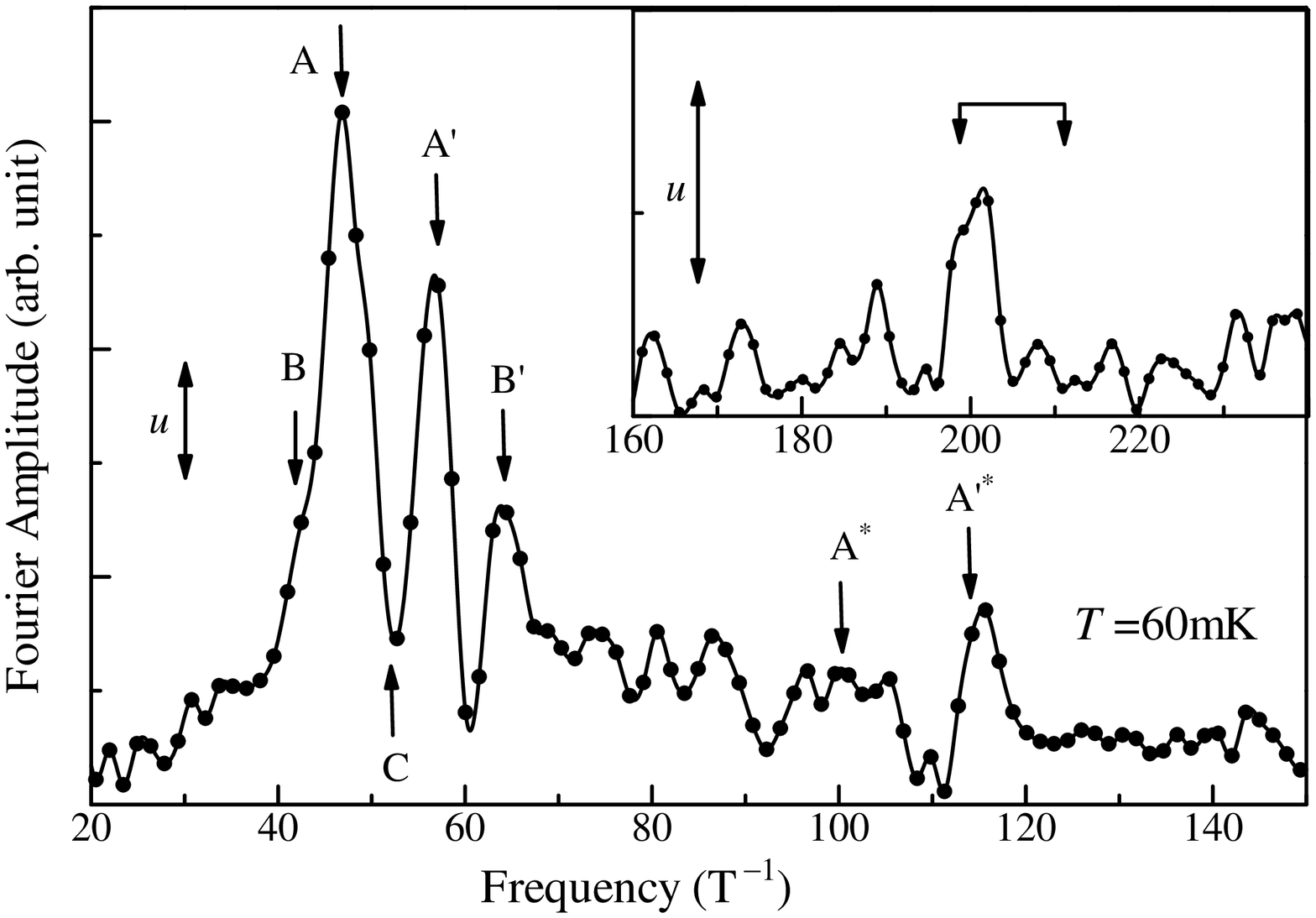}
\caption{Fourier spectrum of the AB-type oscillation in the
magnetoresistance at 60mK. The main peak splits into four (B, A,
A', B') and also the 2nd harmonic splits into two (A$^*$, A'$^*$).
(The other two are within the noise level.) The inset shows peak
structures around four times frequency of the main peak structure
as indicated by arrows. The scales of FT amplitude are shown by
the arrows marked as $u$ in the respective figures.}\label{ft1}
\end{figure}

The observed oscillations are nearly periodic in $B_{\rm ext}$.
The SdH oscillation is not visible in this field range due to the
restriction of the space for free cyclotron motion. The FT
amplitude of the extracted oscillation is displayed in
Fig.\ref{ft1}. The magnetic field range is taken from $-$0.5T to
0.5T. At the main peak around 50T$^{-1}$, clear splitting into
three sub-peaks, marked as A, A' and B', is observed. Sub-peak A
has a shoulder structure at lower frequency marked as B. From
these four splitting frequency positions, the position of the main
peak structure is determined to be 52T$^{-1}$, where a sharp dip
marked as C is located. If we measure the oscillation frequency
with the magnetic flux on the circle with radius $a/2$, position C
just corresponds to the flux quantum $h/e$, confirming that the
peak structure originates from the AB effect around single
anti-dots through the modulation in the density of states.

As noted above, the selection of functional form for background
subtraction does not affect the characteristics shown in
Fig.\ref{ft1} as long as the fitting function is slowly varying
versus $B_{\rm ext}$. We have even performed direct FT without
subtraction. Though the background variation results in
enhancement of unphysical low frequency component, the splitting
of the main peak structure still exists in the result, confirming
the above result.

Another characteristic structure in Fig.\ref{ft1} is
a dip-peak structure around 110T$^{-1}$, which corresponds to
about twice the frequency of the main peak,
and originates from the orbits with the winding number $n$=2.
The signal-to-noise ratio is not high enough to resolve detailed peak structure,
nevertheless we may view it as a dip and two side peaks as indicated by arrows.
Though there is no remarkable structure at the positions at around $n=$3,
a surprisingly clear peak is observed at the position for $n=$4
as shown in the inset of Fig.\ref{ft1}.

\begin{figure}
\includegraphics[width=0.8\linewidth,clip]{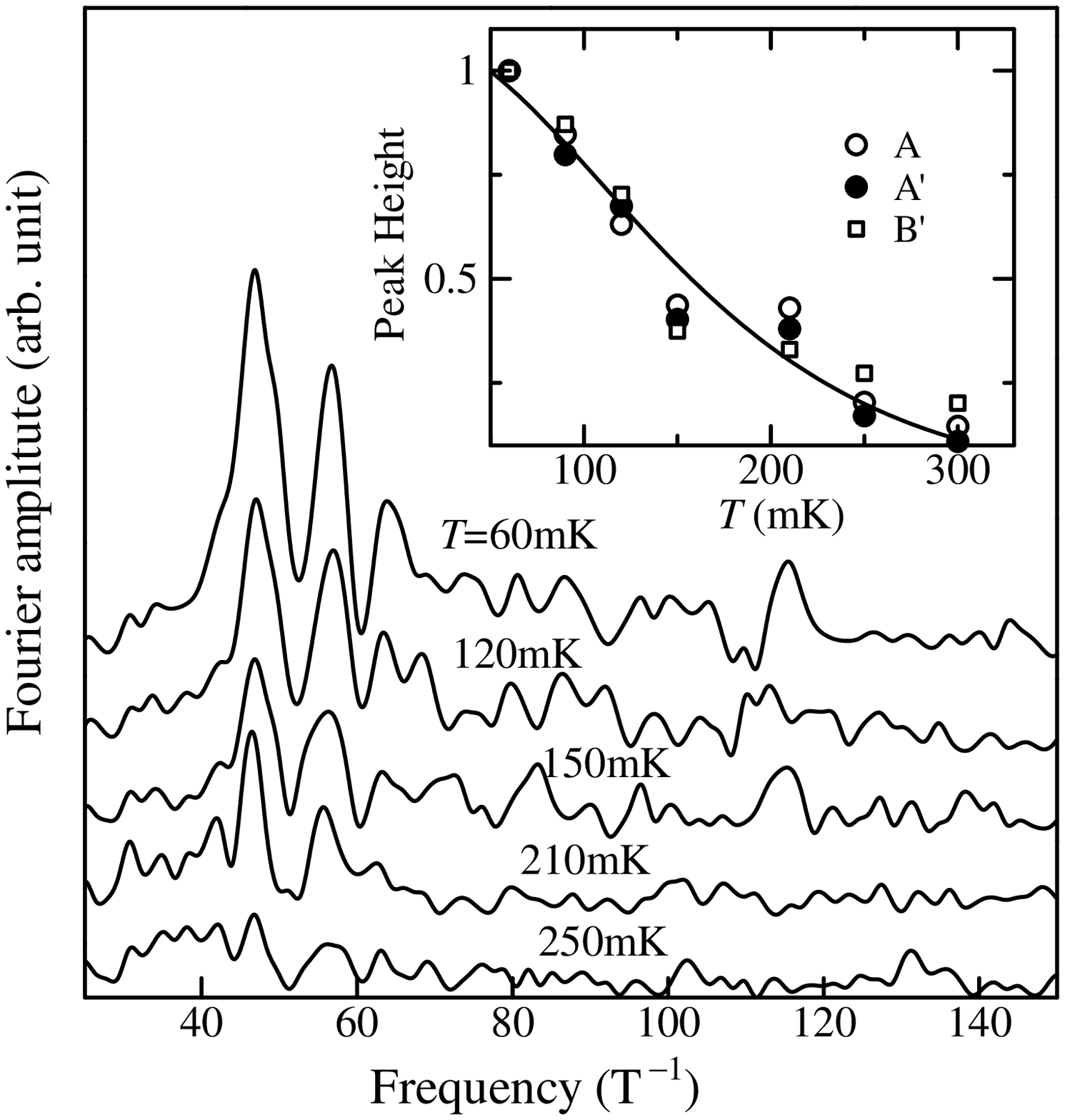}
\caption{FT results at temperatures from 300mK down to 60mK. The
data are offset for clarity. The inset shows the peak heights
normalized at 60mK for peaks A, A' and B'. The solid line in the
inset is the fit to the Dingle function (\ref{eq_dingle}). }
\label{tdep_fig}
\end{figure}

Figure \ref{tdep_fig} displays the temperature dependence of the
FT spectrum. The three peaks (A, A' and B') show almost identical
temperature dependence as seen in the inset of Fig.\ref{tdep_fig},
which can be fitted by the Dingle function,
\begin{equation}
f(T)\propto\frac{\kappa T}{\sinh\kappa T},\quad
\kappa\equiv\frac{2\pi^2k_{\rm B}}{\Delta E},
\label{eq_dingle}
\end{equation}
where $T$ is the temperature and $\Delta E$ the energy difference
between the peaks in the density of states modulation caused by
the AB effect. The fitting gives the value of $\Delta E$ as
86$\mu$eV. This is in similar order with other measurement in 2D
electron systems\cite{Iye}. The agreement in the temperature
dependence manifests the three peaks have the same physical
origin. There, we stressed the nature of the ensemble average in
AB-type oscillations in ADLs, as opposed to the ordinary AB effect
in single ring structure. The AB-type oscillation manifests the
oscillatory fine structure of the density of state, and the
temperature dependence of the oscillation amplitude can be
explained by thermal broadening of the energy levels of the
orbits.

We attribute the origin of the splitting to shifts in additional
phases expressed in Eqs.\ref{ph1} and \ref{ph2}. In the simplest
approximation the oscillatory resistance $\Delta R$ is written as
the sum of four terms:
\begin{equation}
\Delta R = \cos(n(\theta_{\rm AB}+ \Delta \theta_{\rm B}))+
\cos(n(\theta_{\rm AB}- \Delta \theta_{\rm B})) \\
+ \cos(n(\theta_{\rm AB}+ \Delta \theta_k))+ \cos(n(\theta_{\rm
AB}- \Delta \theta_k)), \label{sim_eq}
\end{equation}
where $\theta_{\rm AB}$ is the AB phase given by $2\pi B_{\rm
ext}\cdot \pi r^2 /(h/e)$. Because the phases in Eqs.\ref{ph1} and
\ref{ph2} are not linear in $B_{\rm ext}$, each term in
Eq.\ref{sim_eq} is not sinusoidal oscillation and hence the FT
spectrum strongly depends on the region of transformation. By the
same token, the FT lineshape for higher harmonics ($n>1$) is not
simply $n$-times enlargement in frequency axis of the main peak
structure.

\begin{figure}
\includegraphics[width=0.6\linewidth]{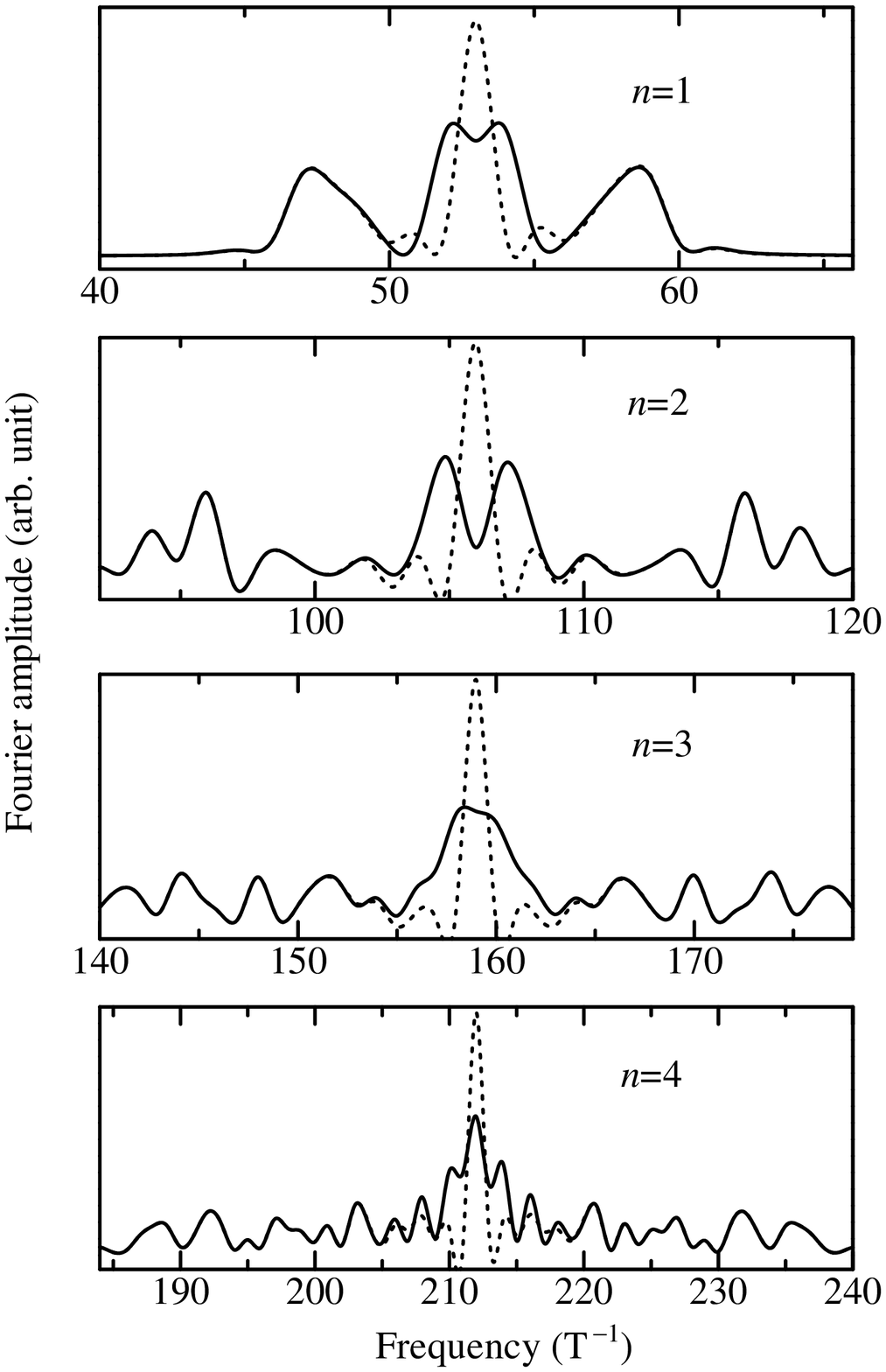}
\caption{Solid lines are the FT spectra of the function
(\ref{sim_eq}) for the winding number $n=$ 1,2,3,4. The magnetic
field region is taken as $-0.5$T$<B<$0.5T and the parameter
$\Delta k_{\rm F}$ is taken from SdH measurement as
4.1$\times$10$^{7}$m$^{-1}$. The scale of abscissa is modified
with $n$ to show the entire peak structures. Dotted lines are the
results when Berry's phase $\Delta\theta_{\rm B}$ is set to zero.
Significant difference between solid and dotted lines around the
center while they are almost identical in surrounding regions. }
\label{fftsim_fig}
\end{figure}

In order for a qualitative comparison with experiment, we
calculate the FT spectra of Eq.\ref{sim_eq} in the magnetic field
ranging from $-$0.5 to 0.5T with $\Delta k_{\rm
F}$=4.1$\times$10$^{7}$m$^{-1}$, which is the value measured from
the SdH oscillation, and $r$ is taken to be $a/2$. We adopt 0.55T
for $B_{\rm eff}$, which is reported for 2DHSs with conditions
close to the present ones\cite{Lu}. The results are shown in
Fig.\ref{fftsim_fig}. Despite the crude approximation, the solid
lines in Fig.\ref{fftsim_fig} bears striking resemblance to the
experiment. For $n$=1, clear splitting into 4 sub-peaks is
observed. With increasing $n$, the distribution of the peaks
increases but the variation of the center peak structure is not
monotonic. For $n$=2, there is a sharp dip at the center while
only a broad peak exists for $n=3$. When $n$ is increased to 4,
the peak distribution still broadens though the center peak
resharpens. These changes in $n=$ 3 and 4 explain the
disappearance and reappearance of peak structure in the
experiment.

As shown by the dotted lines in Fig.\ref{fftsim_fig}, when there
is no contribution of Berry's phase, the characteristics of the
center peak are completely different. No splitting appears even
for $n$ =2; Rounding and sharpening for $n$ =3 and 4 do not appear
either. These results hence support the interpretation that the
splitting of FT peaks is due to the phases in Eqs.\ref{ph1} and
\ref{ph2} arising from the spin-orbit coupling, added to the AB
phase. Equation \ref{sim_eq} is too crude an approximation for
further quantitative discussion. Especially, in ADLs, orbits
encircling anti-dots may vary their sizes with $B_{\rm ext}$,
which is smaller effect than Berry's phase and results in simple
shift of peak positions. Shift of peak positions, {\it e.g.}, for
$n=4$ may be explained with this effect.

It should be remarked that the mean free path of the present 2DHG
is about 600nm, which is much shorter than even the perimeter of a
closed trace encircling a single anti-dot. This fact clearly
emphasizes that the mean free path is essentially different from
the phase coherence length, and also from the circumference of
cyclotron orbit. The former difference has been repeatedly
mentioned in literatures\cite{Imry} while the latter manifests
that the major origin of the scattering is impurity potential,
which allows free space for ballistic cyclotron motions.

Another experimental fact that supports our interpretation is that
the peak splitting in the FT spectra disappears in the AB-type
oscillation in high fields, where the spin is fixed perpendicular
to the plane and neither variation in Berry's phase nor in $\Delta
k_{\rm F}$ occurs. This phenomenon will be reported in a separate
paper.

In conclusion, we have observed spin-orbit Berry's phase
through splitting of Fourier transform peaks in the Aharonov-Bohm type
oscillation of weak field magnetoresistance
in an anti-dot lattice of a two-dimensional hole system.
A simple simulation with adopting parameters obtained
from independent measurements shows fair qualitative
agreement with the experimental observations.

\begin{acknowledgments} The authors would like to thank K. Suzuki for
help during this experiment. This work is supported by a
Grant-in-Aid for Scientific Research from the Ministry of
Education, Culture, Sports, Science, and Technology of Japan and
also supported by Special Coordination Funds for Promoting Science
and Technology.
\end{acknowledgments}


\begin{thebibliography}{99}

\bibitem{Loss} D. Loss, P. Goldbart and A. V. Balatsky, Phys. Rev. Lett.
{\bf 65}, 1655 (1990).
\bibitem{Aronov} A. G. Aronov and Y. B. Lyanda-Geller, Phys. Rev. Lett.
{\bf 70}, 343 (1993).
\bibitem{Stern} A. Stern, Phys. Rev. Lett. {\bf 68}, 1022 (1992).
\bibitem{Qian} T. Z. Qian and Z. B. Su, Phys. Rev. Lett. {\bf 72}, 2311 (1994).
\bibitem{Meir} Y. Meir, Y. Gefen and O. Entin-Wohlman, Phys. Rev. Lett. {\bf 63}, 798 (1989).
\bibitem{Ando2} T. Ando, J. Phys. Soc. Jpn. {\bf 54}, 1528 (1985).
\bibitem{Ekenberg} U. Ekenberg, and M. Altarelli, Phys. Rev. B {\bf 30}, 3569 (1984).
\bibitem{Eisenstein} J. P. Eisenstein, H. L. Stormer, V. Narayanamurti, A. C. Gossard, and W. Wiegmann,
Phys. Rev. Lett. {\bf 53}, 2579 (1984).
\bibitem{Lu} J. P. Lu, J. B. Yau, S. P. Shukla, and M. Shayegan, Phys. Rev. Lett. {\bf 81}, 1282 (1998).
\bibitem{Iye2} Y. Iye, E. E. Mendez, W. I. Wang, and L. Esaki, Phys. Rev. B {\bf 33}, 5854 (1986).
\bibitem{Mopurgo} A. F. Mopurgo, J. P. Heida, T. M. Klapwijk, B. J. van Wees
and G. Borghs, Phys. Rev. Lett. {\bf 80}, 1050 (1998).
\bibitem{Yau} J.-B. Yau, E. P. De Poortere and M. Shayegan,
Phys. Rev. Lett. {\bf 88}, 146801 (2002).
\bibitem{Yang} M. J. Yang, C. H. Yang, and Y. B. Lyanda-Geller, Europhys. Lett. {\bf 66}, 826 (1998).
\bibitem{Weiss} D. Weiss, K. Richter, A. Menschig, R. Bergmann, H. Schweizer, K. von Klitzing, and G. Weimann,
Phys. Rev. Lett. {\bf 70}, 4118 (1993).
\bibitem{Ando} See for review {\it e.g.} T. Ando , S. Uryu , S. Ishizaka and T. Nakanishi,
Chaos, Solitons \& Fractals, {\bf 8}, 1057 (1997).
\bibitem{Nihey} F. Nihey and K. Nakamura, Physica B {\bf 184}
(1993) 398.
\bibitem{Iye} Y. Iye, M. Ueki, A. Endo, and S. Katsumoto, J. Phys. Soc. Jpn. {\bf 73}, 3370 (2004).
\bibitem{Imry} {\it e.g.} see Y. Imry, {\it Introduction to Mesoscopic Physics} 2nd ed. (Oxford, 2001).
\end{thebibliography}
\end{document}